\documentclass[a4paper,11pt]{article}
\usepackage{graphicx}
\usepackage{dcolumn}
\usepackage{bm}
\usepackage{epsfig,amsmath}
\usepackage[usenames,dvipsnames]{color}
\usepackage[pagebackref=false, colorlinks=true]{hyperref}
\usepackage{textcomp}
\usepackage{float}
\title{Extremal limits and Ba\~{n}ados-Silk-West effect}
\author{Parthapratim Pradhan\footnote{pppradhan77@gmail.com}  
\\ {\it Department of Physics, Vivekananda Satabarshiki Mahavidyalaya,
West Midnapur 721513, India}}

\date{}

\begin{document}

\maketitle


\begin{abstract}
A fascinating property of extremal Kerr black hole (BH) is that it could be
act as a particle accelerator with infinite high center-of-mass (CM) energy \cite{bsw}. 
In this note, we would like to discuss about such fascinating
result and to point out that this infinite energy at the event horizon comes solely due
\emph{to the singular nature of the extremal limit}. We also show that a
non-extremal Kerr BH can \emph{not} transform into extremal Kerr BH
by the Ba\~{n}ados-Silk-West  mechanism.
Moreover, we discuss about three possible geometries (near extremal,
purely extremal and near horizon of extremal Kerr) of this mechanism.
We  further prove that near extremal geometry and near
horizon geometry, precisely extremal geometry  of  extremal Kerr BHs
are qualitatively different. Near extremal geometry and near horizon
 geometry gives the CM energy is finite, whereas precisely extremal geometry
gives the diverging energy. Thus, we can argue that extremal Kerr BH
and non-extremal Kerr BH are quite distinct objects.
Finally, we show that the CM energy of collisions of particles not only diverges at infinite 
red-shift surface ($r_{+}$) but it could also diverges at the ISCO ($r_{isco}$) or at the 
circular photon orbit ($r_{cpo}$) or at the marginally bound circular orbit ($r_{mbco}$) or at the 
Cauchy horizon i.e. at $r \equiv r_{isco}=r_{cpo}=r_{mbco}=r_{+}=r_{-}=M$.

\end{abstract}
\maketitle

\section{Introduction}

In 2009, Ba\~{n}ados, Silk and West (BSW) \cite{bsw} proposed an
interesting mechanism for extremal rotating chargeless BH is that it could be act as a particle accelerators with
infinite amount of CM energy, when two massive dark matter particles falling from rest into
the infinite red-shift surface of the BH for some critical values of angular momentum of the in-falling particles. 
Whereas, this energy is finite for non-extremal Kerr BH. For non-rotating Schwarzschild BH the CM energy is also 
finite \cite{baushev}.

This proposal soon criticized by several researchers. Particularly in \cite{berti}, the authors showed that there is
an astrophysical bound i.e. maximal spin, back reaction effect and gravitational radiation etc. on that CM energy 
due to the Thorn's bound \cite{thorn} i.e. $\frac{a}{M}=0.998$ ($M$ is the mass,  $a=\frac{J}{M}$ is the spin
parameter and $J$ is the angular momentum of the BH).  Also in \cite{jacob}, the author derived that the CM 
energy in the near extremal situation for rotating Kerr BH is 
$\frac{E_{cm}}{m_{0}}\sim \frac{4.06}{(1-a)^{1/4}}+{\cal O}((1-a)^{1/4})$. 
Lake \cite{lake} computed that the CM energy at the Cauchy horizon of a static RN BH and Kerr BH  which are bounded. 
Grib and Pablov \cite{grib} investigated the CM energy for Kerr BH using multiple scattering method. The collision in the ISCO 
particles was investigated by the Harada et al. \cite{harada} for Kerr BH. 
On the other hand in \cite{mc}, the author suggested that extremal Kerr BHs are neither particle accelerators nor dark matter 
probes.



In the present study, we shall prove that the CM energy would be  diverging \emph{not} only
at the extremal infinite red-shift surface $(r_{+})$.  Also at the direct ISCO [innermost stable circular orbit
$(r_{isco})$] or at the direct CPO [circular photon orbit ($r_{cpo}$)] or at the direct MBCO [marginally bound circular orbit 
($r_{mbco}$)], the CM energy is also diverging i.e.
\begin{eqnarray}
E_{cm}\mid_{{r_{isco}}}= E_{cm}\mid_{{r_{mbco}}}=
 E_{cm}\mid_{{r_{cpo}}}= E_{cm}\mid_{{r_{+}}} = E_{cm}\mid_{{r_{-}}} \rightarrow \infty
\end{eqnarray}
Thus we may conclude that the CM energy of collision of particles at
$r \equiv r_{isco}=r_{cpo}=r_{mbco}=r_{+}= r_{-}=M$  could be arbitrarily large
for extremal Kerr space-time. On the other hand it can be interpreted as if we choose
the different collision point, say ISCO or CPO or MBCO then we get the same diverging CM energy for each cases.

We will also prove that  the \emph{disparity} between precisely
extremal geometry and non- extremal geometry by using the BSW  effect. One can not transform a non-extremal Kerr
space-time to extremal Kerr space-time via this process. This can be manifested  by computing the
the CM energy and it  is diverging for  extremal Kerr space-time  while the CM energy
is finite for non-extremal space-time. Moreover we also suggest that this diverging energy for \emph{precisely} extremal 
kerr space-time comes solely due to the \emph{singular nature of the extremal limit}, whereas in \cite{bsw} the author 
suggested that this diverging collision energy comes solely due to the gravitational acceleration.

It has long been known that the extremal limit is singular and it is also discontinuous \cite{ddr,cjr,pp1,pp2} has been 
examined from various aspect. In our earlier work \cite{pp2}, we have shown that geodetically an extremal space-time and 
non-extremal space-time are quite distinct objects. For example, in  case of precisely extremal space-time the direct ISCO
which lies on the event horizon which  coincides with the  principal null geodesic generator of
the horizon while the non-extremal space-time do not possess such types of feature.

There are numerous key features which have been present in the extremal space-time
while they are completely absent in the non-extremal space-time. For instance, the
Wald's \cite{wald1} bifurcation $S^{2}$ present in the non-extremal space-time while the extremal space-time
do not possess such $S^{2}$. There are three regions (Region I, Region II and Region III)
in the Carter-Penrose diagram of non-extremal space-times while the extremal
space-times consists of only two regions (Region I and Region III) \cite{c1,c2}. This implies
that Region II is completely absent in extremal geometry while non-extremal geometry
do possesses. This could be interpreted as purely geometric discontinuous nature of the extremal space-time and 
non-extremal space-time due to the lack of the ``Region II''.

Indeed it is true that extremal space-time don't have any trapped surfaces
inside the event horizon while the non-extremal space-time filled up with the 
trapped surfaces \cite{penrose65,pp1,pp2}.

The fact that the proper radial distances between any two points
in the extremal space-time always diverges while the proper distances
between any two  points are finite in the non-extremal situation \cite{ddr}.
Another interesting feature of  extremal geometry is that both
surface gravity and Hawking temperature are zero while they are
non-zero in the non-extremal geometry.

It is also known \cite{grib} that both coordinate time interval $\Delta t$  as well as
proper time interval $\Delta \tau$  diverges for extremal BH  while they are finite in
the non-extremal regime. This is another way to prove the discontinuity between extremal geometry
and non-extremal geometry. It was also mentioned there that the angle $\Delta \phi$ of the
in-falling particle is found to be diverge for extremal Kerr BH while they are finite
for the non-extremal Kerr BH.

In an earlier work \cite{pp3}, we have proved that for spherically symmetric extremal string BH 
could be act as a particle accelerator with diverging  energy at $r \equiv r_{isco}=r_{cpo}=r_{mbco}=r_{+}=2M$.
The main motivation in the present work comes from this work and from Bardeen et al. \cite{bpt}.

\section{Review of ISCOs in  Extremal Kerr BH:}

Let us consider the Kerr metric in Boyer-Lindquist coordinates \cite{bpt} ,
\begin{eqnarray}
ds^2 &=& -\frac{\Delta}{\rho^2} \, \left[dt-a\sin^2\theta d\phi \right]^2+\frac{\sin^2\theta}{\rho^2} \,
\left[(r^2+a^2) \,d\phi-a dt\right]^2
+\rho^2 \, \left[\frac{dr^2}{\Delta}+d\theta^2\right] ~.\label{nkm}
\end{eqnarray}
where
\begin{eqnarray}
a&\equiv&\frac{J}{M},\, \rho^2 \equiv r^2+a^2\cos^2\theta \nonumber\\
\Delta &\equiv& r^2-2Mr+a^2\equiv(r-r_{+})(r-r_{-})
\end{eqnarray}
$M$, $a$ are the mass and angular momentum per unit mass or Kerr parameter respectively.  It may
be noted that  we have used geometrized units through out this work, where $G=c=1=M$.

Now the radius of the event horizon (infinite red-shift surface) and Cauchy horizon are
\begin{eqnarray}
r_{\pm}=1\pm\sqrt{1-a^2}
\end{eqnarray}
which are the roots of the equation $\Delta=0$. The extremal case defined as when $r_{+}=r_{-}$ or $a=M=1$ or $J=M^2=1$ .


The angular velocity of the horizons are
\begin{eqnarray}
 {\Omega}_{\pm} &=& \frac{a}{r_{\pm}^2+a^2} =\frac{a}{2r_{\pm}} ~. \label{omega}
\end{eqnarray}
The ergo-sphere is occur at
\begin{eqnarray}
 r \equiv r_{ergo} &=& 1+\sqrt{1-a^2\cos^2\theta}  ~. \label{ergosp}
\end{eqnarray}

Now let us consider $x^{\mu}(\tau)$ represents the trajectory of the moving particles. $\tau$ is the
proper time of the moving particles. We restrict ourselves the geodesic motion of the particles confined
on the equatorial plane i.e. $u^{\theta}=0$ or $\theta=\pi/2$. Thus the equatorial time-like geodesics for Kerr space-times are
\begin{eqnarray}
  u^{t} &=& \frac{dt}{d\tau}=\frac{1}{\Delta}[(r^2+a^2+\frac{2a^2}{r})E-\frac{2a}{r}L] \\
  u^{r} &=&\frac{dr}{d\tau} =\pm \sqrt{E^{2}+\frac{2}{r^3}(aE-L)^2+\frac{a^2E^2-L^2}{r^2}+\frac{\Delta}{r^2}
  \sigma} \label{eff}\\
  u^{\theta} &=& \frac{d\theta}{d\tau} = 0 \\
  u^{\phi} &=& \frac{d\phi}{d\tau}=\frac{1}{\Delta} [(1-\frac{2}{r})L+\frac{2a}{r}E]~.\label{utur}
\end{eqnarray}
where $\sigma=-1$ for time-like geodesics and $\sigma=0$ for null geodesics. The quantities $E$ and $L$ are
the specific energy and the angular momentum of the particles respectively. The
radial equation for the massive particles moving along the geodesics is described in terms of
effective potential \cite{wald} as :
\begin{eqnarray}
 \frac{1}{2} (u^r)^2+{\cal V}_{eff}(r) &=& 0
\end{eqnarray}
Thus one obtains  ${\cal V}_{eff}(r)$ as :

\begin{eqnarray}
 {\cal V}_{eff}(r) &=& -\frac{1}{r}+\frac{L^2-a^2(E^2-1)}{2r^2}-\frac{(L-aE)^2}{r^3}-\frac{E^2-1}{2}
\end{eqnarray}
The circular orbit of the particle is defined by
\begin{eqnarray}
 {\cal V}_{eff}(r) &=& 0
\end{eqnarray}
and
\begin{eqnarray}
 \frac{d{\cal V}_{eff}(r)}{dr} &=& 0
\end{eqnarray}

Thus we may obtain the energy and angular momentum of the test particle associated with this circular
motions evaluated at $r=r_{0}$ are given by

\begin{eqnarray}
E_{0} &=& \frac{r_{0}^\frac{3}{2}-2r_{0}^\frac{1}{2}\pm a}{r_{0}^\frac{3}{4}\sqrt{r_{0}^\frac{3}{2}-3r_{0}^\frac{1}{2}\pm
2a}} \\
L_{0} &=& \pm \frac{\left(r_{0}^2\mp 2ar_{0}^\frac{1}{2}+a^2\right)}{r_{0}^\frac{3}{4}\sqrt{r_{0}^\frac{3}{2}-3r_{0}^\frac{1}{2}
\pm 2a}}~.\label{angn}
\end{eqnarray}

The upper (lower) sign holds for direct (retrograde) orbits. The innermost stable circular
orbit (ISCO) equation can be obtained by solving the second derivative of the effective potential:
\begin{eqnarray}
 \frac{d^2{\cal V}_{eff}(r)}{dr^2} &=& 0
\end{eqnarray}
Thus one may obtain the ISCO equation for Kerr BH is:
\begin{eqnarray}
 r^2-6r\mp 8a\sqrt{r}-3a^2 &=& 0
\end{eqnarray}
The solution of the equation gives the radius of ISCO for non-extremal Kerr space-times \cite{bpt} are:

\begin{eqnarray}
r_{isco} &=& 3+z_{2}\mp \sqrt{(3-z_{1})(3+z_{1}+2z_{2})}
\end{eqnarray}
where
\begin{eqnarray}
z_{1} &=& 1+(1-a^2)^{1/3} [(1-a)^{1/3}+(1+a)^{1/3}] \\
z_{2} &=& (3a^2+z_{1}^2)
\end{eqnarray}

The circular photon orbit occurs at \cite{bpt}
\begin{eqnarray}
r_{cpo} &=& 2 \{1+\cos[\frac{2}{3}\cos^{-1}(\pm a)]\}
\end{eqnarray}
and
the radius of the marginally bound circular orbit is given by
\begin{eqnarray}
r_{mbco} &=& 2\mp a+2\sqrt{(1\mp a)}
\end{eqnarray}

For extremal BH the direct ISCO occurs at $r_{isco} =M=1$, the direct photon orbit  is at $r_{cpo}=M=1$
and the direct MBCO occurs at  $r_{mbco}=M=1$. Thus three radii
coincides with the horizon \cite{bpt} i.e.
\begin{eqnarray}
r_{isco}=r_{cpo}=r_{mbco}=r_{+}=r_{-}=M=1 
\end{eqnarray}

\section{BSW Effect:}
Now we are going to  compute  the energy in the CM frame for
the collision of two neutral particles coming from infinity with $\frac{E_{1}}{m_{0}}=\frac{E_{2}}{m_{0}}=1$
and approaching the BH with different angular
momenta $\ell_{1}=L_{1}/M$ and $\ell_{2}=L_{2}/M$. The CM energy is
derived by using the formula \cite{bsw} which is valid in both flat
and curved space-time given by
\begin{eqnarray}
\left(\frac{E_{cm}}{\sqrt{2m_{0}}}\right)^{2} &=&  1-g_{\mu\nu}u^{\mu}_{1}u^{\nu}_{2}~.\label{cm}
\end{eqnarray}
where $u^{\mu}_{1}$ and $u^{\nu}_{2}$ are the 4-velocities of the two particles, which can be
determine from the following  equation(\ref{utur}).   Therefore one can compute the CM energy
using the formula (\ref{cm}) \cite{bsw}
\begin{eqnarray}
\left(\frac{E_{cm}}{\sqrt{2}m_{0}}\right)^{2} &=&  \frac{1}{r(r^2-2r+a^2)}[2a^2(r+1)-2a(\ell_{1} +\ell_{2})
-\ell_{1}\ell_{2}(r-2)+2r^2(r-1)- \nonumber \\[4mm] &&
\sqrt{2(a-\ell_{2})^2-{\ell_{2}}^2r+2r^2}\sqrt{2(a-\ell_{1})^2-\ell_{1}^2r +2r^2}]   ~.\label{cm1}
\end{eqnarray}
Thus the limiting value of CM energy $E_{cm}$ at the horizon $r=r_{+}$ is given by
\begin{eqnarray}
E_{cm} &=& 2m_{0}\sqrt{1+\frac{(\ell_{1}-\ell_{2})^2}{2r_{-}(\ell_{1}-\ell_{H})(\ell_{2}-\ell_{H})}} ~.\label{cm2}
\end{eqnarray}
The ranges of the angular momentum per unit rest mass for in-falling geodesics are given by
\begin{eqnarray}
-2\left(1+\sqrt{1+a}\right)\le \ell \le 2\left(1+\sqrt{1-a}\right)~.\label{cam}
\end{eqnarray}
Substituting the values of $\ell_{1}=2\left(1+\sqrt{1-a}\right)$ and
 $\ell_{2}=-2\left(1+\sqrt{1+a}\right)$ in (\ref{cm2}), and $\ell_{H}=\frac{2r_{+}}{a}=\frac{2(1+\sqrt{1-a^2})}{a}$,
 we get the CM energy for non-extremal Kerr BH reads as \cite{grib}
\begin{eqnarray}
E_{cm} &=& \frac{2m_{0}}{(1-a^2)^{1/4}}\sqrt{\frac{(1-a^2)+(1+\sqrt{1+a}+\sqrt{1-a})^2}{1+\sqrt{1-a^2}}} ~.\label{cm3}
\end{eqnarray}
This is indeed a finite quantity. When $a=J$ then CM energy reads as
\begin{eqnarray}
E_{cm} &=& \frac{2m_{0}}{(1-J^2)^{1/4}}\sqrt{\frac{(1-J^2)+(1+\sqrt{1+J}+\sqrt{1-J})^2}{1+\sqrt{1-J^2}}} ~.\label{cm4}
\end{eqnarray}
It follows from the above analysis  the CM energy depends on the spin of the BH.
Now if we take the extremal limit $a=J=1$, CM energy diverges. Therefore we have drawn
the following conclusions:
(a) We can not obtain the extremal Kerr space-time  by taking the extremal limit of a non-extremal Kerr space-time.
(b) BSW mechanism  can not transform non-extremal Kerr BH to extremal Kerr BH.
(c) The infinite amount of CM energy for extremal BH solely due to the singular nature
 of the extremal limit. The geometric discontinuity between two space-time gives the diverging
 energy.

Let us now discuss briefly about the BSW effect for
the three possible geometries namely, near extremal geometry, extremal
geometry and near horizon geometry of extremal Kerr space-time.

\textbf{(a) Near Extremal  Geometry:}

It is astro-physically relevant to compute such geometry for the limiting \cite{sch}
behavior of $r_{isco}$, $r_{cpo}$, $r_{mbco}$. Taking $a=1-\chi$; then
\begin{eqnarray}
r_+ &=& 1 + (2\chi)^{1/2}+ O(\chi^{3/2})\,\,,
r_- = 1 - (2\chi)^{1/2}+ O(\chi^{3/2})\nonumber\\
r_{cpo}&=& 1+\sqrt{\frac{8\chi}{3}}+O(\chi^{3/2})\,\, ,
r_{mbco}= 1+2\sqrt{\chi}\nonumber\\
r_{isco}&=& 1+\sqrt[3]{4\chi}\,\, ,
r_{ergo}= 2 \nonumber\\
\ell_{1}&=& 2(1+\sqrt{\chi}) \,\, ,
\ell_{2}=-2(1+\sqrt{2-\chi})
\end{eqnarray}
Now calculating the CM energy for the above cases we get,
\begin{eqnarray}
\left(\frac{E_{cm}}{\sqrt{2}m_{0}}\right)^{2}\mid_{{r=r_{isco}}} &=& \frac{F(r_{isco})}{G(r_{isco})}
\end{eqnarray}
where
\begin{eqnarray}
F(r_{isco}) &=& 2a^2(r_{isco}+1)-2a(\ell_{1} +\ell_{2})
-\ell_{1}\ell_{2}(r_{isco}-2)+2r_{isco}^2(r_{isco}-1)- \nonumber \\[4mm] &&
\sqrt{2(a-\ell_{2})^2-{\ell_{2}}^2r_{isco}+2r_{isco}^2}\sqrt{2(a-\ell_{1})^2
-{\ell_{1}}^2r_{isco} +2r_{isco}^2}\\
G(r_{isco})&=& r_{isco}(r_{isco}^2-2r_{isco}+a^2)
\end{eqnarray}
Now substituting the values  of $r_{isco}\approx 1+\sqrt[3]{4\chi}$,
we have the CM energy:
\begin{eqnarray}
 E_{cm}\mid_{{r_{isco}}} \, \propto \frac{1}{\chi^{1/3}}
\end{eqnarray}
Similarly, we can obtain for $r_{\pm} \approx 1 \pm (2\chi)^{1/2}$, the
CM energy:
\begin{eqnarray}
 E_{cm}\mid_{{r_{\pm}}} \, \propto \frac{1}{\chi^{1/4}}
\end{eqnarray}

Similarly, we get for $r_{mb}\approx 1+2\sqrt{\chi}$,
the CM energy:
\begin{eqnarray}
 E_{cm}\mid_{{r_{mbco}}} \, \propto \frac{1}{\chi^{1/2}}
\end{eqnarray}
and for $r_{cpo} \approx 1+\sqrt{\frac{8\chi}{3}}$,
the CM energy:
\begin{eqnarray}
 E_{cm}\mid_{{r_{cpo}}} \,  \propto \frac{1}{\chi^{1/2}}
\end{eqnarray}
In each cases, we have seen that the CM energy $E_{cm}$
is finite for any generic values of $\ell_{1}$ and $\ell_{2}$.

The interesting phenomenon could be occur at the extremal limit $\chi =0$.
\begin{eqnarray}
 E_{cm}\mid_{{r_{+}}} &=& E_{cm}\mid_{{r_{-}}}=E_{cm}\mid_{{r_{isco}}}= E_{cm}\mid_{{r_{mbco}}}=
 E_{cm}\mid_{{r_{cpo}}} \rightarrow \infty
\end{eqnarray}
Interestingly, the CM energy at each collision point is diverging. In fact this infinite
energy comes solely due to the singular nature of the extremal limit and
due to the purely geometric discontinuity between the two space-time.

It is noteworthy to mentioned here that the expression for CM energy at the ergo-sphere $r_{ergo}=2$ is
found to be
\begin{eqnarray}
\left(\frac{E_{cm}}{\sqrt{2}m_{0}}\right)^{2} &=& \frac{F(r_{ergo})}{G(r_{ergo})}
\end{eqnarray}
where
\begin{eqnarray}
F(r_{ergo}) &=& 8-2(1-\chi)(\ell_{1} +\ell_{2})+6(1-2\chi)- \nonumber \\[4mm] &&
\sqrt{2(1-\chi-\ell_{2})^2-2{\ell_{2}}^2+8} \sqrt{2(1-\chi-\ell_{1})^2
-2{\ell_{1}}^2 +8}\\
G(r_{ergo})&=& 2(1-2\chi)
\end{eqnarray}
Now we may turn to the  precisely extremal case to see what happens there.

\textbf{(b) Precisely Extremal  Geometry:}

For the extremal cases $a=M=1$ and at the extremal horizon $r=M=1$, we obtain
the CM energy \cite{bsw}:
\begin{eqnarray}
 E_{cm}\mid_{{r_{+}=M=1}} &=& \sqrt{2}m_{0}\sqrt{\frac{\ell_{2}-2}{\ell_{1}-2}+\frac{\ell_{1}-2}{\ell_{2}-2}}
\end{eqnarray}
This is also finite for  any generic values of $\ell_{1}$ and $\ell_{2}$.
Here the CM energy strongly depends on the critical values of the
angular momentum. When $\ell_{1}=2$ and $\ell_{2}=2$, the CM energy
is diverging i.e.
\begin{eqnarray}
 E_{cm}\rightarrow \infty
\end{eqnarray}
On the other hand using equations (\ref{cm3}) and (\ref{cm4}), at the
extremal limit $a=M=J=1$, it can be directly  seen that
\begin{eqnarray}
 E_{cm}\rightarrow \infty
\end{eqnarray}
Thus it proves that one can not obtain the extremal geometry
via taking the extremal limit of a non-extremal/near-extremal
geometry. The fact that CM energy is diverging in the extremal
limit suggests that one should regard non-extremal and precisely
extremal BHs as qualitatively distinct objects. Thus
one can not turn a non-extremal Kerr BH into a extremal Kerr
BH via limiting procedure because the extremal limit is
singular. Alternatively, we can say that the diverging energy
comes due to the singular nature of the extremal limit.

Whereas at the ergo-sphere the CM energy is found to be
\begin{eqnarray}
E_{cm}\mid_{{r_{ergo}=2}} &=& m_{0}\sqrt{14+4\sqrt{2}-\sqrt{17+8\sqrt{2}}}
\end{eqnarray}
It is  indeed finite for both near-extremal case and extremal case.
For our record, we also discuss the near horizon geometry
of extremal Kerr (NHEK) BH.

\textbf{(c) Near Horizon Geometry of Extremal Kerr BH:}

In \cite{gala}, the author computed the CM energy
in the near horizon setup for Kerr-Newman BH. We apply
the same formalism here  for calculating the CM energy for
the near horizon geometry of extremal Kerr background. Then we
 compare the results obtained in three possible geometries, near
extremal geometry, extremal geometry and near horizon of extremal
geometry.

In this geometry, the new coordinates are
\begin{eqnarray}
r \rightarrow 1+\epsilon r_{0}r, t \rightarrow \frac{tr_{0}}{\epsilon},
\phi \rightarrow \phi +\frac{ta}{\epsilon r_{0}}
\end{eqnarray}
where $r_{0}^2=1+a^2$, in such a way that in the new frame the horizon
is situated at $r=0$. Then one takes the limit $\epsilon \rightarrow 0$,
which yields
\begin{eqnarray}
ds^2 &=& {\rho_{0}}^2 (-r^2dt^2+\frac{dr^2}{r^2}+d\theta^2)+\frac{(1+a^2)^2\sin^2\theta}{{\rho_{0}}^2}
(d\phi+\frac{2a}{1+a^2} rdt)^{2}  ~.\label{nhkm}
\end{eqnarray}
where ${\rho_{0}}^2=1+a^2\cos^2\theta$.

This is also a vacuum solution of the Einstein's equations. Now restrict our
attention to the equatorial  geodesics, one obtains
\begin{eqnarray}
  u^{t} &=& \frac{dt}{d\tau}= \frac{1}{r^2}-\frac{2a\ell}{(1+a^2)r} \\
  u^{r} &=&\frac{dr}{d\tau} =\pm \sqrt{1-\frac{4a\ell}{(1+a^2)}r-
  \frac{[\ell^2 (1-4a^2)+(1+a^2)^2]}{(1+a^2)^2}r^2}\\
  u^{\theta} &=& \frac{d\theta}{d\tau} = 0 \\
  u^{\phi} &=& \frac{d\phi}{d\tau}=
  -\frac{(1-4a^2)\ell}{(1+a^2)^2}-\frac{2a}{(1+a^2)r}
  ~.\label{nhkp}
\end{eqnarray}

Now deriving the CM energy near the horizon $r\rightarrow 0$
is given by
\begin{eqnarray}
E_{cm} &=& m_{0}\sqrt{\frac{16a^2+(\ell_{1}-\ell_{2})^2}{4a^2}}
~.\label{cm5}
\end{eqnarray}
This is in-fact CM energy of near horizon of non-extremal Kerr
BH. This is indeed finite for any values of $\ell_{1}$ and
$\ell_{2}$.

In the extremal case $a=M=1$, the CM energy is found to be
\begin{eqnarray}
E_{cm} &=& m_{0}\sqrt{\frac{16+(\ell_{1}-\ell_{2})^2}{4}}
~.\label{cm6}
\end{eqnarray}

It is indeed finite for any generic values of $\ell_{1}$ and $\ell_{2}$.

\section{Discussion}
Thus we have argued that extremal Kerr BHs are fundamentally different class from non-extremal
Kerr BHs because they have diverging CM energy. This is why we can claim that the non-extremal Kerr BH 
can not be transformed into extremal one by the BSW mechanism. The other reasons are  the topology of two
space-times are drastically different. Thus extremal Kerr geometry can not be obtained as
limits of non-extremal Kerr geometry. Since the \emph{extremal limit itself is a singular }. Therefore the 
divergence energy generates due to this fact. The other aspect we have studied that
the diverging energy in the CM frame due to the singular nature of the
extremal limit. We also showed that the near extremal geometry, precisely extremal geometry and near
horizon geometry of  extremal Kerr BHs are qualitatively
different. Near extremal geometry and near horizon geometry gives
the CM energy is finite whereas purely extremal geometry gives the diverging energy.

Another way we can explained as extremal BH and non-extremal
Kerr BHs are physically quite distinct objects and it is
impossible to transform the former to the latter by the BSW mechanism.
The another feature of this work is that for extremal BH three
radii  namely ISCO, CPO and MBCO coalesces to both the  horizons namely event horizon \cite{bpt}  and Cauchy horizon 
thus we may vary the collision point but they all gives the identical result (diverging energy). 
Thus in conclusion, for extremal Kerr BH it is shown that the CM energy of collision at 
$r \equiv r_{isco}=r_{cpo}=r_{mbco}=r_{+}=r_{-}=M=1$ is arbitrarily large. This diverging CM energy for the
colliding particles can only be attained when the BH is precisely extremal and only at infinite  coordinate time 
as well as infinite proper time. To sum up, we suggests the infinite amount of collision energy at different collision 
point comes solely \emph{due to the singular nature of the extremal limit}



\bibliography{apssamp}

\end{document}